\documentstyle[pra,aps,multicol,eqsecnum,epsfig]{revtex}

\newcommand\be{\begin{eqnarray}}
\newcommand\ee{\end{eqnarray}}
\newcommand\ba{\begin{array}}
\newcommand\ea{\end{array}}
\def\r{\rangle}
\def\l{\langle}
\def\T{{\rm Tr}}

\begin{document}

\title{Quantum homogenization}
\author{
M. Ziman${}^{1}$, P. \v{S}telmachovi\v{c}${}^{1}$,
V. Bu\v{z}ek${}^{1,2}$,
M. Hillery${}^{1,3}$,
V. Scarani${}^{4}$, and
N. Gisin${}^{4}$
}
\address{
${}^{1}$ Research Center for Quantum Information,
Slovak Academy of Sciences,
D\'ubravsk\'a cesta 9, 842 28 Bratislava, Slovakia \\
${}^{2}$Faculty of Informatics, Masaryk University, Botanick\'a 68a,
602 00 Brno, Czech Republic
\\
${}^{3}$Department of Physics, Hunter College of CUNY,
 695, Park Avenue, New York, NY 10021, U.S.A.
\\
${}^{4}$Groupe de
Physique Appliqu\'{e}e, Universit\'{e} de Gen\`{e}ve, 20 rue de
l'Ecole de M\'{e}decine, 1211 Gen\`{e}ve 4, Switzerland
\\
}

\date{23 October 2001}

\maketitle

\begin{abstract}
We design a {\em universal} quantum homogenizer, which is a quantum
machine that takes as an input
a system qubit initially in the state $\rho$
and a set of $N$ reservoir qubits
initially prepared in the same  state $\xi$.
In the homogenizer the system qubit sequentially
interacts with the reservoir qubits via the {\em partial
swap} transformation.
The homogenizer realizes, in the limit sense, the transformation
such that at the output each qubit
is in an arbitratily small
neighbourhood of the state $\xi$
irrespective of the initial states of the system and the
reservoir qubits. This means that the system qubit undergoes an
evolution that has a fixed point, which is the reservoir state
$\xi$. We also study approximate homogenization when the reservoir
is composed of a finite set of identically prepared qubits. The
homogenizer allows us to understand various aspects of the
dynamics of open systems interacting with environments in
non-equilibrium states. In particular, the reversibility {\em vs}
or irreversibility of the dynamics of the open system is directly
linked to specific (classical) information about the order in
which the reservoir qubits interacted with the system qubit. This
aspect of the homogenizer leads to a model of a quantum safe with
a classical combination.We analyze in detail how entanglement
between the reservoir and the system is created during the process
of quantum homogenization. We show that the information about the
initial state of the system qubit is stored in the entanglement
between the homogenized qubits.

\end{abstract}

\begin{multicols}{2}

\section{Introduction}
\label{sec1}

When a system interacts with a reservoir which is in thermal
equilibrium then after some time the system is thermalized - it
relaxes towards the thermal equilibrium.  This implies that the
information about the original state of the system is (irreversibly)
``lost''  and its new state
is determined exclusively by
the parameters (temperature) of the reservoir. If the reservoir is
composed of a large number, $N$, of physical objects
of the same physical type as the system itself, then the
thermalization process can be understood as homogenization:
out of $N$ objects (the reservoir) prepared
in the same thermal state and a single system in an
arbitrary state, we obtain $N+1$ objects in the
same thermal state. This intuitive picture is based on certain assumptions
about the interaction between the system and the reservoir, about the
physical nature of the reservoir itself and the concept of the thermal
equilibrium. This picture is at the heart of the model of
blackbody radiation, which triggered the birth of quantum theory
in the seminal work of Planck.
In addition, this same picture is very important
in understanding many processes in quantum physics as well as
the fundamental concept of the irreversibility \cite{Peres,Davies}.

In this paper we  present a rigorous analysis of the above picture
within the framework of quantum information theory. Specifically,
we will consider a system, $S$, represented by a single qubit
initially prepared in the unknown state $\varrho_S^{(0)}$, and a
reservoir, $R$, composed of $N$ qubits all prepared in the state
$\xi$, which is arbitrary but same for all qubits. We will
enumerate the qubits of the reservoir and denote the state of the
$k$-th qubit as $\xi_k$ \cite{footnote1}. From the definition of
the reservoir it follows that initially $\xi_k=\xi$ for all $k$,
so the state of the reservoir is described by the density matrix
$\xi^{\otimes N}$.

\begin{figure}
\includegraphics[width=8cm]{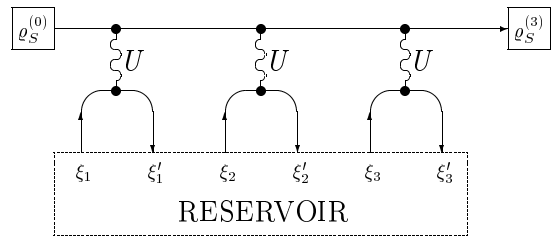}
\caption{The scenario of homogenization with just three
reservoir qubits involved.}
\label{fig1}
\end{figure}

Let $U$ be a unitary operator representing the interaction between
a system qubit and one of the reservoir qubits. In addition,
let us assume that at each time step the system qubit interacts with
just a single qubit from the reservoir (see Fig.~\ref{fig1}).
Moreover, the system qubit can
interact with each of the reservoir qubits at most once. After the
interaction with the $1$-st reservoir qubit the system is changed
according to the following rule (which is a completely-positive map)
\be
\label{1.1}
\varrho_S^{(1)} =\T_1
\left[U\varrho_S^{(0)}\otimes\xi_1 U^\dagger\right].
\ee

Let us repeat the interaction $N$ times, that is, via a sequence
of interactions the system qubit interacts with $N$ reservoir
qubits all prepared in the state $\xi$. The final state of the
system is then described by the density operator \be \label{1.2}
\varrho_S^{(N)} &=& \T_R\left[U_N\dots
U_1\left(\varrho_S^{(0)}\otimes \xi^{\otimes N}\right)
U_1^\dagger\dots U_N^\dagger\right] \ee where
$U_k:=U\otimes(\bigotimes_{j\ne k}\openone_j)$ describes the
interaction between the $k$-th qubit of the reservoir and the
system qubit. This model of homogenization is very similar to the
{\em collision model} since the system becomes homogenized via a
sequence of individual interactions  with the reservoir qubits.
The interactions are assumed to be localized in time (i.e., they
act like ellastic collisions) \cite{Alicki}.

Our aim is to investigate   possible maps
induced by the transformation (\ref{1.2})
and  describe the process of homogenization.
Homogenization means that due to the
interaction $U$ the states of the qubits in reservoir change only little
while
after $N$ interactions the system's state become close to the initial state
of the reservoir qubits. Formally,
\be
\label{1.3}
\forall  N\geq N_{\delta}&
\dots\dots& D(\varrho_S^{(N)},\xi)\le\delta\, ; \\
\label{1.4}
\forall k,  1\leq k \leq N & \dots\dots &
D(\xi_{k}^\prime,\xi)\le\delta ,
\ee
where $D(.,.)$ denotes some distance (e.g., a trace norm)
between the states, $\delta>0$ is a
small parameter which is chosen a priori to the determine the
degree of the homogeneity and
$\xi_{k}^\prime:=\T_S [U\varrho_S^{(k-1)}\otimes\xi U^\dagger]$
is the state of the $k$-th reservoir qubit after the interaction with
the system qubit.

\begin{figure}
\includegraphics[width=8cm]{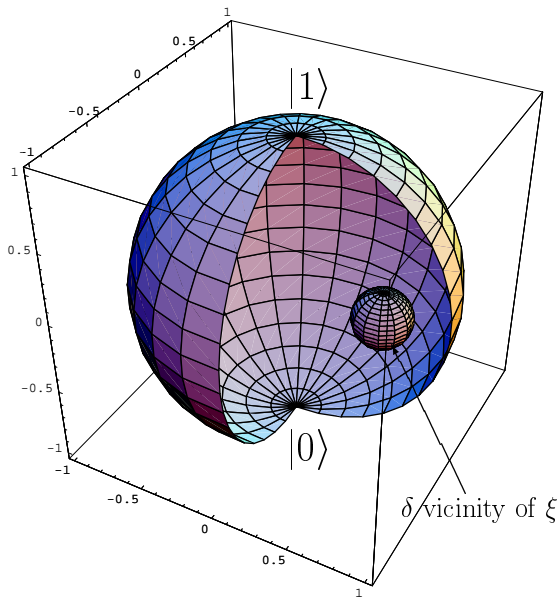}
\caption{
The $\delta$ neighborhood of the reservoir's state $\xi$ inside the Bloch
sphere. After
$N$ interactions between the system and the reservoir
the states  of all reservoir  qubits  and
the system qubit are contained
within this $\delta$-sphere.}
\label{fig2}
\end{figure}

The conditions (\ref{1.3}) and (\ref{1.4}) can be represented using a
geometrical picture. The Bloch sphere of unit radius
is a representation of the state space of a
spin-1/2 (qubit) system. The initial state $\rho$ of the system qubit and
the reservoir state $\xi$ are represented by two (distinct) points
of the Bloch sphere. We can image another sphere of the radius
$\delta$ centered at the point representing the reservoir state $\xi$
(in what follows we will call this sphere the $\delta$-sphere).
The task is to ``shrink'' the original Bloch sphere representing the
(unknown) initial state space of the system qubit into the $\delta$-sphere.
So we start with $N$ reservoir qubits in the state $\xi$ and the system
qubit in an arbitrary state $\rho$ and we end up
with $N+1$ qubits within the $\delta$-sphere centered at the point
representing the original reservoir state $\xi$ (see Fig.~\ref{fig2}).

We note that homogenization is closely related to {\em
thermalization}. There are however two main differences: in
thermalization, (i) the state $\xi$ of the reservoir qubits is not
completely unknown, but is a thermal state, that is, a state
diagonal in a {\em given} basis (interpreted as the basis of the
eigenstates of a one-qubit Hamiltonian); and (ii) the number of
qubits in the reservoir is considered to be infinite for any
practical purpose. Thermalization is studied in Ref.
\cite{Scarani}.

Our paper is organized as follows:
in Section \ref{sec2} we show that  quantum homogenization
can be realized with the help of
a {\em partial swap} operation. In Section \ref{sec3}
we show that the partial swap for qubits generates
 a contractive map on the system qubit with the
fixed point being the initial state of the reservoir. This ensures
the required convergence of the homogenization process [see
Eqs.~(\ref{1.3}) and (\ref{1.4})]. The uniqueness of the
partial-swap operation is proved in Section \ref{sec4}. In Section
\ref{sec5}
 we estimate the fidelity of the approximate
homogenization map as a function of the number $N$ of
reservoir qubits  and the
parameter $\delta$ (the precision of the homogenization),
while in the Section \ref{sec6} we will analyze how the reservoir
qubits become entangled as a consequence of their interaction
with the system qubit. In the
final Section \ref{sec7} of the paper we
address possible applications of the homogenization map.

\section{Partial-swap operation}
\label{sec2}

Let us start with the definition of the so-called {\it swap}
operation $S$ acting on the Hilbert space of two qubits which is
given by relation \cite{Nielsen2000}
\be
\label{2.1}
S |\psi\r\otimes|\phi\r=|\phi\r\otimes|\psi\r.
\ee
With this transformation
\be
\label{2.2}
S\,\varrho^{(0)}\otimes\xi\, S^{\dagger}\,=\,\xi\otimes
\varrho^{(0)}\; , 
\ee
after just a single interaction, the state of
the system $S$ is equal to the state $\xi$ of the reservoir qubit;
and the interacting qubit from the reservoir is left in the
initial state of system. This means the condition (\ref{1.3}) is
fulfilled, while the condition (\ref{1.4}) is not --- since recall
that we want it to hold for all $\varrho^{(0)}$.

In order to fulfill both conditions (\ref{1.3}) and (\ref{1.4}) we
have to find some unitary transformation which is ``close'' to the
identity on the reservoir qubit, while it performs a {\em partial}
swap operation, so that the system qubit at the output is closer
to the reservoir state $\xi$ than before the interaction. The swap
operator is Hermitian and therefore we can define the  unitary {\em
partial swap} operation
\be
\label{2.3}
P(\eta)=\cos\eta\,\openone \,+\,i\,\sin \eta\, S
\ee
that serves our purposes. In what follows we denote $\sin
 \eta=s$ and $\cos  \eta=c$.

In the process of homogenization, the system qubit interacts
sequentially with one of the $N$ qubits of the reservoir through
the transformation $P( \eta)$. After the each interaction,
the system
qubit becomes entangled with the qubit of the reservoir with which
it interacted (for more details on the issue of entanglement
see Sec.~\ref{sec6})
The states of the system qubit and of the reservoir
qubit are obtained by partial traces. Specifically, after
the first interaction the system qubit is
in the state described by the density operator
\be
\label{2.4}
\varrho_S^{(1)}=c^2\varrho_S^{(0)}+s^2\xi+ics[\xi,\varrho_S^{(0)}],
\ee
while the first reservoir qubit is now in the state
\be
\label{2.5}
\xi_{1}^\prime=s^2\varrho_S^{(0)}+c^2\xi+ics[\varrho_S^{(0)},\xi].
\ee
We can recursively apply  the partial-swap transformation and
after the interaction with
the $n$-th reservoir qubit, we have
\be
\label{2.6}
\varrho_S^{(n)}=c^2\varrho_S^{(n-1)}+s^2\xi+ics[\xi,\varrho_S^{(n-1)}]\,
,
\ee
as the expression for the density operator of the system
qubit, while the $n$-th reservoir qubit is in the state
\be
\label{2.7}
\xi_{n}^\prime=
s^2\varrho_S^{(n-1)}+c^2\xi+ics[\varrho_S^{(n-1)},\xi]\, .
\ee
Since we are interested only in those terms in expressions
(\ref{2.6}) and (\ref{2.7}) that are proportional to the operator
$\xi$ we can rewrite the above equations in the form
\be
\label{2.8}
\varrho_S^{(n)}=s^2\sum_{k=0}^{n-1}c^{2k}\xi+\rho_{rest}^{(n)}=
(1-c^{2n})\xi+\rho_{rest}^{(n)} \, ,
\ee and
\be
\label{2.9}
\xi_{n}^\prime=
s^2(1-c^{2(n-1)})\xi +\xi_{n,rest}\, .
\ee
In the next section, we are
going to show that $\rho_{rest}^{(n)}$ converges monotonically to
the null operator as $n\rightarrow\infty$. In this case, obviously
$\varrho_S^{(n)}\rightarrow\xi$, so the condition (\ref{1.3}) is
fulfilled if the number of qubits $N$ is large enough. In addition,
as $n$ increases, $\xi'_n$ becomes more and more similar to $\xi$,
since the commutator in (\ref{2.7}) goes to zero; in other words,
\be
\label{2.10} D(\xi_{n}^\prime,\xi)\le
D(\xi_{n-1}^\prime,\xi)\, .
\ee
Therefore, condition (\ref{1.4}) will be
fulfilled for all $k$ if and only if it is fulfilled for $k=1$.
This gives us a restriction on the parameter $ \eta$ that enters
the partial swap; this restriction will be studied in Section
\ref{sec5}.

\section{Homogenization is a contractive map}
\label{sec3}

In this section we want to show that $\varrho^{(N)}_S\to\xi$
monotonically, for all parameters $\eta\neq 0$. This means, in
particular, that condition (\ref{1.3}) does not put any constraint
on $\eta$. To show this convergence, we use the {\it Banach
theorem} \cite{Simon&Reed80} that concerns the fixed point of a
contractive transformation. Let $\cal S$ be a space with a
distance function $D(\varrho,\xi)$, then the transformation $T$ is
called {\em contractive} if it fulfills  the inequality
$D(T[\varrho],T[\xi])\le kD(\varrho,\xi)$ with $0\le k<1$ for all
$\varrho,\xi\in\cal S$. A fixed point of the transformation $T$ is
an element of $\cal S$ for which $T[\xi]=\xi$. The Banach theorem
states that a contractive map has a unique fixed point
\cite{note0}, and that the iteration of the map converges to it,
i.e. $T^N[\varrho]\to\xi$ for each $\varrho\in\cal S$. We note
that contractive transformations within the context of quantum
information processing have been recently
 discussed also in Ref.~\cite{Raginski01}.

In our case $\cal S$ is the set of physical states, i.e. the set
of all density matrices of a single qubit. The map, $T$, that we
are considering is defined by $\varrho^{(0)}_S\to
T[\varrho^{(0)}_S]=\varrho^{(1)}_S$. We must show that the map is
contractive, and that $\xi$ is a fixed point of the map.

We begin by finding the super-operator induced by the
transformation $U$ in the left-right form, i.e., as a linear
operator acting on the space of trace-class operators ${\cal
T}({\cal H})$ (see Ref.~\cite{Ruskai01}). We choose the operators
$\frac{1}{2}\openone,\sigma_x,\sigma_y,\sigma_z$ (where
$\sigma_{n}$ are the Pauli matrices) as a basis for ${\cal
T}({\cal H})$, where $\cal H$ represents the Hilbert space of a
qubit. In this case an arbitrary density operator of a qubit
can be written as \be \label{3.1}
\varrho=\frac{1}{2}\openone+\vec{w}\cdot\vec{\sigma} \ee where
$|\vec{w}|\le 1/2$. We can write a state that is an element of
${\cal T}({\cal H})$ in a vector form, i.e.
$\varrho=(1,w_x,w_y,w_z)$. Let
$\xi=\frac{1}{2}\openone+\vec{t}\cdot\vec{\sigma}=(1,t_x,t_y,t_z)$
be the state of the qubit in the reservoir. After the first
interaction $P$ with the first
 reservoir qubit the system qubit
evolves according to Eq.(\ref{2.6}) with $n=1$. This
transformation can be described as \be
\varrho^{(0)}_S\to\varrho^{(1)}_S&=&
s^2\xi+c^2\varrho^{(0)}_S+ics[\xi,\varrho^{(0)}_S] \nonumber
\\
&=&\frac{1}{2}\openone+(s^2\vec{t}+c^2\vec{w})\cdot\vec{\sigma}+ics
[\vec{t}\cdot\vec{\sigma},\vec{w}\cdot\vec{\sigma}]
\nonumber
\\
&=&\frac{1}{2}\openone+[s^2\vec{t}+c^2\vec{w}-2cs
(\vec{t}\times\vec{w})]\cdot\vec{\sigma}
\nonumber
\\
&=& \frac{1}{2}\openone + \vec{w}^{~\prime} \cdot\vec{\sigma},
\label{3.2}
\ee
where we used the identity
$\sigma_k\sigma_l=\delta_{kl}\openone+i\varepsilon_{jkl}\sigma_j$,
and
\be
w^\prime_j=s^2 t_j+(c^2\delta_{jl}-2cs\, \varepsilon_{jkl}t_k)w_l \, ,
\label{3.3}
\ee
with $j=x,y,z$.
Now we can express the transformation
$\varrho^{(0)}_S\to\varrho^{(1)}_S$
as
\begin{equation}
\label{3.4}
\left(
\ba{c} 1\\ w_x^\prime \\ w_y^\prime \\ w_z^\prime\ea
\right) =
\left(\ba{cccc}
1 & 0 & 0 & 0\\
s^2 t_x & c^2 & 2cst_z & -2cst_y\\
s^2 t_y & -2cst_z & c^2 & 2cst_x\\
s^2 t_z & 2cst_y & -2cst_x & c^2\\
\ea\right)\left(\ba{c}1 \\ w_x \\ w_y \\ w_z \ea\right),
\end{equation}
or more formally, as
$\varrho^{(1)}_S=T\varrho^{(0)}_S$, where $T$ is the matrix
representing the super-operator acting on the linear space
${\cal T}({\cal H})$.
If we express the matrix $T$ as
\be
\label{3.5}
T=\left(
\ba{cc}
1 & \vec{0}^T\\
s^2\vec{t} & {\bf T} \ea \right) \, , \ee then it is easy to check
that in our case ${\bf T}\vec{t}=c^2 \vec{t}$. This implies that
the state $\xi$ is a {\em fixed} point of the map under
consideration, i.e. $T\xi=\xi$.  The
system state after $n$-th iteration then reads 
\be 
\varrho_S^{(n)}
&=&\frac{1}{2}\openone +\left[\sum_{j=0}^{n-1} s^2 {\bf
T}^j\vec{t}+ {\bf T}^n\vec{w}\right]\cdot\vec{\sigma} \nonumber
\\
& =& \frac{1}{2}\openone +\left[s^2\sum_{j=0}^{n-1} c^{2j}\vec{t}+
{\bf T}^n\vec{w}\right]\cdot\vec{\sigma} \,  \nonumber
\\ &=&
\label{3.7} \frac{1}{2}\openone +\left[(1-c^{2n})\,\vec{t}+ {\bf
T}^n\vec{w}\right]\cdot\vec{\sigma} \, ,
\ee 
where for the last
equality we summed the geometric sum $\sum_{j=0}^{n-1}
(c^2)^j=(1-c^{2n})/(1-c^2)$. Of course, $c^{2n}\to 0$ unless
$c=\cos\eta= 1$.  Numerically one can check
that ${\bf T}^n\to{\bf O}$, where ${\bf O}$ represents the zero
operator. Thus $\varrho^{(n)}_S\to\xi$. In what follows we prove
this convergence for all values of the parameter $\eta$.

To prove that the map $T$ is contractive, we must define a
distance function on $\cal{S}$. Let us introduce the trace
distance $D(\varrho,\omega)=\T|\varrho-\omega|$ and the vectors
$\omega=(1,v_x,v_y,v_z)$ and $\vec{r}=\vec{w}-\vec{v}$.
For a qubit we have \be \label{3.8}
D(\varrho,\omega)=\T|(\vec{w}-\vec{v})\cdot\vec{\sigma}|=\T|\vec{r}
\cdot \vec{\sigma}|\,=\,2|\vec{r}|\ee since the eigenvalues of the
operator $\vec{r}\cdot\vec{\sigma}$ are given by
$\lambda_{\pm}=\pm |\vec{r}|$. In order to find
 the contraction parameter $k$ for our transformation $T$ we proceed
as follows. From the Eqs.~(\ref{3.5}) and (\ref{3.8})  we obtain
\be
D(T[\varrho],T[\omega])=\T
|\vec{r}^{~\prime}\cdot\vec{\sigma}|\,=\,2|\vec{r}^{~\prime}|,
\ee where $\vec{r}^{~\prime} =\vec{w}^{~\prime}-\vec{v}^{~\prime}=
s^2\vec{t}+{\bf T}\vec{w}-s^2\vec{t} -{\bf T}\vec{v}={\bf
T}(\vec{w}-\vec{v})={\bf T}\vec{r}
=c^2\vec{r}-2cs\vec{t}\times\vec{r}$. Since
$|\vec{t}|^2\le 1/4$ and
$|\vec{r}^{~\prime}|^2=c^4|\vec{r}|^2+4c^2s^2|\vec{t}\times\vec{r}|^2=
|\vec{r}|^2c^2(c^2+4s^2|\vec{t}|^2\sin^2\beta)$, where
$\beta\le\pi$ is the angle between the vectors $\vec{t}$ and
$\vec{r}$, we find that the contraction coefficient $k=c$. This
last equality
 is due to the fact
that $|\vec{r}^{~\prime}|\le|\vec{r}|c$. If  $c=\cos \eta<1$
then the map $T$ is contractive and the convergence to the fixed
point $\xi$ is assured.

\section{Uniqueness of the partial-swap operation}
\label{sec4}
In what follows we will
discuss the question of the choice of the unitary transformation,  ${ U}$,
that describes the interaction between a system from the reservoir and the
initial system undergoing the homogenization process. If both the system
and the reservoir state are the same, the interaction should not affect
either qubit, and this should be true no matter what the state of the
system and reservoir qubit are. This implies that the unitary operator
must satisfy the following two conditions:
\begin{eqnarray}
\label{c1}
{\rm Tr}_1 \left ( { U} \rho \otimes \rho  { U}^{\dagger} \right ) & = & \rho \; , \\
\label{c2}
{\rm Tr}_S \left ( { U} \rho \otimes \rho { U}^{\dagger} \right ) & = & \rho \; ,
\end{eqnarray}
for any single-qubit state, $\rho$.
Let us first discuss the case of pure states. If $\rho$ represents a pure
state then the condition (\ref{c1}) says that $ {U} \rho \otimes \rho {
U}^{\dagger} = \rho \otimes \xi_1 $ where $\xi_1$ needs to be determined.
However from the second condition (\ref{c2}) it follows that
$  { U} \rho \otimes \rho  {U}^{\dagger} = \xi_2 \otimes \rho $
where $\xi_2$ is unknown. Putting the last two results together we obtain that
\begin{eqnarray}
\label{c3}
 {U} \rho \otimes \rho  { U}^{\dagger} = \rho \otimes \rho \; ,
\end{eqnarray}
for any $\rho$ representing a pure state. From here it follows that the unitary
transformation ${ U}$ acting on the joint Hilbert space ${\cal H}^2 = {\cal
H} \otimes {\cal H}$ must be of the form
\begin{eqnarray}
\label{c4}
{ U}: |\psi \rangle \otimes | \psi \rangle \rightarrow e^{i \varphi }|\psi \rangle \otimes | \psi \rangle \; ,
\end{eqnarray}
where the parameter $\varphi $ is independent of the state $|\psi \rangle$.
Therefore,
the action of the unitary
transformation is fixed on the symmetric subspace of ${\cal H}^2$ up to a
phase factor $ e^{i\varphi }$. Neither of the two conditions (\ref{c1}) and
(\ref{c2}) nor the condition (\ref{c4}) tell us anything about
the action of the unitary
transformation ${ U}$  on the antisymmetric subspace of ${\cal H}^2$.
This means that action
of ${ U}$ on the antisymmetric subspace is arbitrary.  However, in
the case of qubits the antisymmetric subspace is one dimensional,
and we  can proceed further.
Because the antisymmetric subspace is one dimensional and
invariant under the action of
the unitary transformation ${ U}$, we have
\begin{eqnarray}
\label{c5}
{U} ( | \psi \rangle | \psi^{\perp}
\rangle -| \psi^{\perp}  \rangle | \psi \rangle ) =
e^{i\theta}\left( | \psi \rangle | \psi^{\perp} \rangle -
| \psi^{\perp}  \rangle | \psi \rangle \right)\; ,
\end{eqnarray}
where $\theta$ is a constant depending on ${U}$. Now the transformation
${U}$ is given by the equations (\ref{c4}) and (\ref{c5}) up to two
constants $\varphi$ and $\theta$. What we would now like to show is
that these conditions require that $U$ be a partial swap operator
up to a global phase factor.  This phase factor has no physical
consequences.  If we define the unitary operator $U^{\prime}$ to be
\begin{eqnarray*}
{U}' = \exp^{i(-\theta - \varphi)/2} { U} \; ,
\end{eqnarray*}
then equations (\ref{c4}) and (\ref{c5}) give us
\begin{eqnarray*}
 | \psi \rangle | \psi \rangle
&\stackrel{U^\prime}{\rightarrow }&
e^{i (\varphi  - \theta) / 2} | \psi \rangle | \psi \rangle
\\
| \psi \rangle | \psi^{\perp} \rangle -| \psi^{\perp}  \rangle | \psi \rangle )
&\stackrel{U^\prime}{\rightarrow }&
e^{i (\theta-\varphi ) / 2 }
\left(| \psi
\rangle | \psi^{\perp} \rangle -| \psi^{\perp}  \rangle | \psi \rangle
\right)
 \; .
\end{eqnarray*}
Comparing these equations to Eq. (\ref{2.3}), we see that $U^{\prime}$
is just the {\em partial-swap} operator with $\eta = ( \varphi -
\theta )/2$. We can, therefore, conclude
that in the case of qubits, the partial swap is the {\em only}
possible operator that satisfies the conditions of homogenization (\ref{c1})
and (\ref{c2}). The partial-swap uniquely determines yet another universal
quantum machine \cite{machines}: {\em the universal quantum homogenizer}.

\section{Approximate homogenization}
\label{sec5}
In what follows we will analyze homogenization not as the
limit of the infinite number of interactions, but as an
approximate process after a finite number of steps. Let us suppose
that the parameter $\delta$ from Eqs. (\ref{1.3}) and (\ref{1.4})
is fixed. This parameter characterizes our approximation. We will
use the partial-swap evolution for the description of the
homogenization.

In the first step we give a condition on the parameter $\eta$ of
the partial swap (\ref{2.3}). For our  map $T$, we have that
$D(\varrho_S^{(N)},\xi)\le D(\varrho_S^{(N-1)},\xi)\le
D(\varrho_S^{(0)},\xi)$. On the other hand from Eq. (\ref{2.10})
we know that $D(\xi_{N}^\prime,\xi)\le D(\xi^\prime_{N-1}, \xi)$.
As we have discussed earlier, we can adjust the parameter $\eta$
so that the condition $D(\xi_{1}^\prime,\xi)\le\delta$ is
fulfilled. Obviously, the distance $D(\xi_{1}^\prime,\xi)$ depends
on the initial state of the system, $\varrho_S^{(0)}$, and on
$\eta$. Therefore we have to determine the maximum value of
$\eta$, for which the distance is less than or equal to $\delta$,
independent (the universality condition) of the initial states of
the system and reservoir. For a qubit the maximum value of trace
distance is achieved for $\vec{w}=-\vec{t}$, corresponding to the
situation in which the states are pure and mutually orthogonal.
The argument for this can be easily  seen from a geometric
representation of a qubit.  In this case
\be \label{4.3} D(\xi^\prime_{1} ,\xi)=2 s^2
\T|\vec{t}\cdot\vec{\sigma}|=2 s^2 \ee since for a pure state
$|\vec{t}|=\frac{1}{2}$. From Eq. (\ref{4.3}) we get the simple
relation 
\be 
\label{rel} 
\sin\eta\,\le\,\sqrt{\delta/2} \; . 
\ee

The second step is to determine the minimum number of interactions,
$N$, that ensures for an arbitrary initial state of the system that
the final state is in a sphere of radius $\delta$ around the
reservoir state $\xi$. The worst case, i.e. when the number of
necessary iterations is maximal, is intuitively the case when
$D(\varrho_S^{(0)},\xi)$ is maximal. In section \ref{sec3} we
proved the convergence of the system state to $\xi$ for any
$\eta \neq 0$.  Therefore we are sure that such an $N$ exists.
As was just discussed in the previous paragraph, the distance
$D(\varrho_S^{(0)},\xi)$ is maximal, when the two states are pure
 and mutually orthogonal.
Moreover,
our transformation $T$ doesn't change the
commutation relation, which is initially equal to zero,
 i.e.  $[\varrho_S^{(N)},\xi]=0$, for all $N$.
Introducing  $\vec{w}=-\vec{t}$ for the commuting states we obtain
\be \label{4.1} \varrho_S^{(N)}=\frac{1}{2}\openone
+(1-2c^{2N})\vec{t}\cdot\vec{\sigma}\, , \ee and for the distance
we find \be \label{4.2}
D(\varrho_S^{(N)},\xi)=\T|(\vec{w}^{~\prime}-\vec{t})\cdot\vec{\sigma}|=
2c^{2N}\T|\vec{t}\cdot\vec{\sigma}|\, . \ee This distance is
maximal if we fix  $N$ and maximize over all $\varrho_S^{(0)}$ and
$\xi$.  Again, since
$|\vec{t}|=\frac{1}{2}$ for pure states, we obtain the distance
$D(\varrho_S^{(N)},\xi)=2c^{2N}=2(\cos\eta)^{2N}$. If the parameters
$\eta$ and $s$ in 
the experession (\ref{rel})  are such that 
$\sin\eta = \sqrt{\delta/2}$ then we can find the lower bound 
$N_\delta$ on the
number of reservoir qubits which are necessary to achieve the
homogenization with a required fidelity
\be 
\label{4.4} 
N\,\geq N_{\delta} =
\,\frac{\ln \delta/2}{\ln(1-\delta/2)}\, .
\ee 
Both bounds on the parameters $\eta$ and $N$ are completely
determined by the parameter $\delta$.  After
performing $N$ iterations, $N+1$ qubits are in  states belonging
to the $\delta$ neighborhood of the initial state of the
reservoir, no matter what the states $\xi$ and $\varrho_S^{(0)}$
were.

We see that
if we fix the number of reservoir's qubits $N$,
then the other two parameters are
determined by the relations (\ref{4.4}) and (\ref{rel}).

\section{Entanglement via homogenization}
\label{sec6}
In spite all the progress in the understanding of the nature
of quantum entanglement there are still open questions that
have to be answered. In particular, a problem which
waits for a thorough illumination is the nature of
multiparticle entanglement
\cite{Thapliyal1999}. There are several aspects of quantum
multiparticle correlations that have been investigated recently.
One example is the investigation
of intrinsic $N$-party entanglement (i.e, generalizations
of the GHZ state \cite{GHZ1990}). Another is the realization
that in contrast to classical correlations, entanglement cannot
freely be shared among many objects.

Coffman {\it et al.} \cite{Coffman2000} have recently studied a
set of three qubits, and have proved that the sum of the
entanglement (measured in terms of the tangle) between the qubits
$1$ and $2$ and the qubits $1$ and $3$ is less than or equal to
the entanglement between qubit $1$ and the rest of the system,
i.e. the subsystem $23$. Specifically, let us define the
bi-partite concurrence  \cite{Hill1997} of a two-qubit system in
the state $\varrho_{jk}$ to be: \be \label{definition} C_{jk}
\equiv C(\varrho_{jk}):=
\max\{0,\lambda_1-\lambda_2-\lambda_3-\lambda_4\} \ee where the
$\lambda_i$s are the square roots of the eigenvalues of the matrix
 $R=\varrho_{jk}(\sigma_y\otimes\sigma_y)
(\varrho_{jk})^*(\sigma_y\otimes\sigma_y)$ 
listed indecreasing order.  The tangle is equal to the square of
the concurrence, i.e. $\tau_{jk}=[C_{jk}]^2$. Using this
definition we can express the Coffman-Kundu-Wootters (CKW)
\cite{Coffman2000} inequality as \be \label{8.1} C_{12}^2+C_{13}^2
\leq C^2_{1,(23)} \, . \ee In the same paper they conjectured that
a similar inequality might hold for an arbitrary number, $N$, of
qubits prepared in a pure state. That is, one has \be \label{8.2}
\sum_{k=1;k\neq j}^N C^2_{j,k}\leq C^2_{j,\overline{j}}\, , \ee
where the sum in the left-hand-side is taken over all qubits
except the qubit $j$, while $C^2_{j,\overline{j}}$ debotes the
concurrence between the qubit $j$ and the rest of the system
(denoted as $\overline{j}$).

Several interesting results in the investigation of various
bounds on entanglement in multi-partite systems have been reported
recently. In particular, Wootters \cite{Wootters}
has considered an {\em infinite} collection of qubits arranged in an
open  line,
such that every pair of nearest neighbors is entangled.
In this translationally invariant
{\em entangled chain} the maximum closest-neighbor (bi-partite)
entanglement (measured in the concurrence) is bounded by the value
$1/\sqrt{2}$ (it is not known whether this bound is achievable)
\cite{Wootters}. Later  Koashi {\em et al.}
\cite{Koashi2000} considered
a {\em finite} system of $N$ qubits in which
each pair out of $N(N-1)/2$ possible pairs is  entangled
(the so-called web of entanglement).
It has been proved that the maximum possible bi-partite
concurrence in this case is equal to $2/N$. D\"{u}r \cite{Dur2001}
considered other possible inequalities associated with
variously entangled qubits in multi-partite systems.

Within the context of our investigation
it is very natural to ask, what is the nature of the entanglement created
during the process of homogenization. In this section we will
address several questions related to this issue.
Firstly, we will study the bi-partite entanglement
between the system qubit and the reservoir qubits, and then
we will analyze
entanglement between reservoir qubits which is induced by the interaction
with the system qubit. We will show that CKW bounds are saturated, that
is the $N+1$ qubit state created by a sequence of
partial-swap operations in the homogenization process satisfies the
inequality in Eq.~(\ref{8.2}) as an equality.

\subsection{Bipartite concurrence}
Let us consider the  concurrence  $C^{(n)}_{jk}$ between the
$j-$th and $k-$th qubits (irrespective whether these are reservoir
or system qubits) after the $n-$th interaction, assuming that
initially the system was in the state $\varrho$ and the
reservoir qubits were in the state $\xi$. Without loss of
generality we shall always assume that $j<k$. The value $j=0$
denotes the system qubit and $j=1,2,\dots, N$ denote the qubits of
the reservoir.

The reduced density operator $\varrho^{(n)}_{jk}$
describing the two qubits under consideration
is given by the expression
\be
\label{8.3}
\varrho^{(n)}_{jk}=\T_{\overline{j} \overline{k}}\left(
{\cal U}_n\dots{\cal U}_1[\varrho\otimes\xi^{\otimes N}]\right)\, ,
\ee
with ${\cal U}_l[\sigma]=P_l\sigma P_l^\dagger$, where $P_l$ is the partial
swap operation acting between the system qubit and the $l$-th qubit of
the reservoir [see Eq.(\ref{2.3})].
The line over the indices in the trace formula denote the partial
trace over all subsystems except the ones with the line over them.

Using the definition (\ref{definition}) of the concurrence, it
is trivial to see that $C^{(n)}_{jk}=0$ for $j,k>n$,
that is, the qubits which have not interacted yet are not entangled.
On the other hand, a general expression for the concurrence is
difficult to derive, so we concentrate our attention on a special
case, when the reservoir is initially
in a {\em pure} state $|\xi\r$ while
the system qubit is in an arbitrary state $\varrho$.

Following the homogenization scenario the system qubit after
$(k-1)$ interactions is in the state $\varrho_0^{(k-1)}$,
that can be expressed in the basis
$\{|\xi\r,|\xi^\perp\r\}$ as
\be
\nonumber
\varrho_0^{(k-1)} &=& a_{k-1}|\xi\r\l\xi|
+(1-a_{k-1})|\xi^\perp\r\l\xi^\perp|\\
& &
+b_{k-1}|\xi\r\l\xi^\perp|+b_{k-1}^*|\xi^\perp\r\l\xi|.
\ee
After we apply the $k$-th partial swap operation between the system
and the $k$-th reservoir qubit we find the bi-partite density operator
in the matrix form (in the given basis)
\begin{equation}
\nonumber
\varrho_{0k}^{(k)}=\left(
\begin{array}{cccc}
a_{k-1} & cb_{k-1} & isb_{k-1} & 0 \\
b_{k-1}^*c & (1-a_{k-1})c^2 & isc(1-a_{k-1}) & 0 \\
-isb_{k-1}^* & -isc(1-a_{k-1}) & s^2(1-a_{k-1}) & 0 \\
0 & 0 & 0 & 0
\end{array}
\right)
\end{equation}
where $c=\cos \eta, s=\sin \eta$.

The matrix, $R$, constructed from  $\varrho^{(k)}_{0k}$
has only one nonzero eigenvalue $4c^2s^2(1-a_{k-1})^2$. This implies for the
concurrence
\be
C_{0k}^{(k)}=2cs(1-a_{k-1}).
\ee
From  Eq.(\ref{2.6}) we find the recurrence formula for the
parameters $a_k$
\be
a_k=a_{k-1}c^2 +s^2=1-c^{2k}(1-a_0),
\ee
from which we obtain
\be
C_{0k}^{(k)}=2csc^{2(k-1)}(1-a_0)
\ee
where $a_{0}:=\l\xi|\varrho|\xi\r$, and $C_{0k}^{(k)}$
is the concurrence
measuring the  entanglement between the system qubit
and $k-$th reservoir qubit just after their join interaction
(i.e., it is supposed that the system qubit has interacted all together
just $k$ times).
We can conclude that the system qubit is entangled with $k$-th reservoir
qubit. On the other hand we can ask whether this entanglement will
persist after the system interacts later with other reservoir qubits.
In order to make the discussion simpler we will study a particular case
when initially the system is in the state $|1\r$ while the reservoir
qubits are in the state $|0\r$. Nevertheless, prior to this task we study
another aspect of multi-partite entanglement within the context
of homogenization. Specifically, we will study how a given qubit
is entangled with the rest of the system.

\subsection{One qubit {\it vs} rest of the system}

In  the case of pure multi-qubit states one can define a measure
of the entanglement between a single qubit and the rest of the
system \cite{Coffman2000} with the help of the determinant of the
density operator of the specific qubit under consideration. In
particular, let us begin the homogenization process with the
system and the reservoir qubits initially in pure states. After
$n$ partial swaps the $j$th qubit is in the state
$\varrho_j^{(n)}=\T_{\overline{j}}\left({\cal U}_n\dots{\cal
U}_1[|\psi\r |\xi\r^{(\otimes N)}]\right)$. The degree of
entanglement between the $j$th qubit and the rest of the system is
given by the expression \cite{Coffman2000} \be
\tau^{(n)}_{j}\equiv \left[C_{j,\overline{j}}\right]^2 :=4{\rm
det}\varrho^{(n)}_j\, \ee where $\tau^{(n)}_{j}$ is the tangle,
which is equal to the square of the corresponding concurrence.

Obviously, for the $j$th qubit of the reservoir, the tangle is
zero until it interacts with the system qubit. After the
interaction its value remains constant, irrespective of the
further evolution of the system qubit during the homogenization
process. This means that
\begin{equation}
\label{tangle1}
\tau^{(n)}_{j}=\left\{
\ba{lcr}
0 &\ {\rm if} &\ n<j\le N\\
4{\rm det}\xi_j^\prime &\ {\rm if} &\ j\le n\le N
\ea\right.
\end{equation}
In order to justify the last equation we note that
all measures of entanglement remain unchanged under local unitary
transformations, and that
all  transformations ${\cal U}_k$ (except the $j$th
one) are local with respect to the partition $j\oplus \overline{j}$
(where $\overline{j}$ denotes all qubits except $j$th).

The tangle between the system qubit and the reservoir
is given by the expression
\be
\tau^{(n)}_{0}=4{\rm det}\varrho_0^{(n)},
\ee
from which it follows
that the shared entanglement between the system qubit and the whole
reservoir depends on the total number
of interactions $n$, unlike
in the case (\ref{tangle1}) of the reservoir qubits.

\subsection{The case $|\psi\r_0=|1\r$ and $|\xi\r_j=|0\r$}
In order to have a deeper insight into the problem of entanglement
induced by the homogenization process, let us consider a specific
initial state of the system and the reservoir:
$|\psi\r_0=|0\r$ and $|\xi\r_j=|1\r$. In this case we find for the
tangle between the system and the rest of the reservoir qubits
after the $n$-th interaction the expression
\be
\label{c_0^n}
\tau^{(n)}_{0}= 4\det\varrho_0^n=4 c^{2n}(1-c^{2n}) ,
\ee
since $\varrho_0^{(n)}=(1-c^{2n})|0\r\l 0|+c^{2n}|1\r\l 1|$
[cf. Eq.~(\ref{2.8})].
It is clear from this expression that as $n\rightarrow\infty$
the degree of entanglement between the system and the reservoir
is monotonically decreasing.
On the other hand, the state of $j$-th qubit after the interaction
with the system qubit is
$\xi_j^\prime=s^2\varrho_0^{(j-1)}+c^2|0\r\l 0|=
(1-s^2 c^{2(j-1)})|0\r\l 0|+s^2 c^{2(j-1)}|1\r\l 1|$
[cf. Eq.~(\ref{2.9})] from which it follows that
\be
\label{c_j^j}
\tau^{(j)}_{j}= 4s^{2}c^{2(j-1)}(1-s^{2}c^{2(j-1)})
\ee
In other words,  after its interaction with the system qubit the
$j$-th qubit is constantly entangled with the rest of the
system. These simple examples  illustrate the more
general conclusions presented in the previous paragraph.

Let us turn our attention to the
bipartite concurrences $C_{jk}^{(n)}$.
With the given initial conditions, we easily find the state
vector describing the whole system after $n$ interactions:
\be
|\Psi\r=
c^n|1\r_0\otimes |0\r^{\otimes N}+\sum_{l=1}^n
|1\r_l\otimes|0\r^{\otimes N_{\overline{l}}}
\left[isc^{l-1}(c+is)^{N-l}\right]
\, .
\ee
We recall
that $N$ is the total number of reservoir qubits, and that
we have assumed that $j<k$.
The state $|0\r^{\otimes N_{\overline{l}}}$ denotes all
qubits except the qubit $l$ in the state $|0\r$.
Tracing over the appropriate
subsystems we find the density matrices for $j<n<k$,
\be
\nonumber
\varrho_{jk}^{(n)}&=&\xi_j^\prime\otimes|0\r\l 0|\, ,\\
\varrho_{0k}^{(n)}&=& \varrho_0^{(j)}\otimes|0\r\l 0|.
\ee
For  $k\le n$ we find
\be
\nonumber
\varrho_{jk}^{(n)}&=&\left[c^{2n}+\sum_{l=1,l\ne k,j}^n s^2c^{2(k-1)}
\right]|00\r\l 00|\\
\nonumber & & +
s^2 c^{2(k-1)}|01\r\l 01|+s^2 c^{2(j-1)}|10\r\l 10|\\
\nonumber & & + s^2 c^{j+k-2}(c+is)^{k-j}|01\r\l 10|\\
& & + s^2 c^{j+k-2}(c-is)^{k-j}|10\r\l 01| \, ,
\ee
and
\be
\nonumber
\varrho_{0k}^{(n)}&=&\sum^n_{l=1,l\ne k}s^2 c^{2(l-1)}|00\r\l 00|\\
\nonumber & & +
c^{2n}|10\r\l 10|+s^2c^{2(k-1)}|01\r\l 01|\\
\nonumber & & +isc^{n+k-1}(c+is)^{n-k}|01\r\l 10|\\
& & -isc^{n+k-1}(c-is)^{n-k}|10\r\l 01| \, ,
\ee
which determines the values of the concurrences.
The corresponding eigenvalues of the matrices $R_{jk}^{(n)}$,
constructed from the density matrices
$\varrho_{jk}^{(n)}$ (in the case $n>k$),
 are
\be
\nonumber
{\rm eig}(R_{jk}^{(n)}) &= &\{4 s^4 c^{2(j+k-2)},0,0,0\}\\
{\rm eig}(R_{0k}^{(n)})& = & \{4 s^2 c^{2(n+k-1)},0,0,0\}.
\ee
The square roots of these eigenvalues are the $\lambda_i$s
in  Eq.~(\ref{definition}). For the concurrences we find
\begin{equation}
\label{8.18}
C_{jk}^{(n)}=\left\{ \ba{lcl} 0 &\ {\rm for}&\ n<k\le N \\
2s^2c^{j+k-2} &\ {\rm for}&\ k\le n\le N \ea \right.  \, ,
\end{equation}
\begin{equation}
C_{0k}^{(n)}=\left\{
\ba{lcl} 0 &\ {\rm for}&\ n<k\le N \\
2sc^{n+k-1} &\ {\rm for}&\ k\le n\le N
\ea
\right. \, .
\label{8.19}
\end{equation}
We see that the concurrence between any two  qubits of
the reservoir is zero until
both of them have interacted with the system qubit.
Then the concurrence rises during the interaction to a new value and
remains constant in the subsequent evolution.
On the other hand the value of the entanglement
between the system qubit and any qubit from the reservoir
becomes nonzero after their joint interaction,
but then it tends back to zero.

This means that the system qubit acts as a ``mediator'' of entanglement
between the reservoir qubits which have never interacted directly.
It is obvious that later the two reservoir qubits interact with
the system qubit, the smaller the degree of their mutual entanglement.
Nevertheless, this value is constant and does not depend on the subsequent
evolution of the system qubit (i.e., it does not depend on the number of interactions $n$).

Once we have derived expressions for the bipartite concurrences, we can
verify the CKW inequality (\ref{8.2}),
which in our notation takes the form
\be
\label{woot}
S_j(n):=\sum_{k=1}^N \left[C_{jk}^{(n)}\right]^2\le \tau_{j}^{(n)}\equiv
\left[C_{j,\overline{j}}^{(n)}\right]^2\, .
\ee

Firstly, let us consider the entanglement of the system
qubit with the reservoir.
Using Eq.~(\ref{8.19}) we can explicitly evaluate the
expression for the left-hand-side of the inequality (\ref{woot})
and we can compare it with the expression
(\ref{c_0^n}) representing the right-hand-side of this
inequality. We find that:
\be
\label{s_0}
S_0(n)=\sum_{k=1}^n [C_{0k}^{(n)}]^2=4c^{2n}(1-c^{2n})=\tau^{(n)}_0 \, ,
\ee
which means that the bound on the bi-partite entanglement
between the system and the reservoir qubits is saturated and
the two sides are equal.

In fact, this property is  also valid for the reservoir qubits.
So, let us consider a qubit $j$ of the reservoir.
In the case $n<j$ all the $C_{jk}^{(n)}$ vanish. That means
$S_j(n)=0=\tau_{j}^{(n)}$.
If $n\ge j$ then
\be
\nonumber
S_j (n)&=&[C^{(n)}_{0j}]^2+\sum_{k=1}^{j-1}[C_{kj}^{(n)}]^2+\sum_{k=j+1}^n
[C_{jk}^{(n)}]^2 \\
\nonumber
&=& 4s^2 c^{2(n+j-1)}+4s^4 c^{2(j-2)}\left(\sum_{k=1}^{j-1}c^{2k}+
\sum_{k=j+1}^{n}c^{2k}\right)\\
\nonumber
&=&4s^2 c^{2(n+j-1)}+4s^4 c^{2(j-2)}(c^2 -s^2 c^{2j}-c^{2(n+1)})\\
&=&4s^2 c^{2(j-1)}(1-s^2 c^{2(j-1)}).
\ee
In the calculation we used the equality
\be
\sum_{k=1}^{j-1}c^{2k}+
\sum_{k=j+1}^n c^{2k}=\sum_{k=0}^n c^{2k}-1-c^{2j}.
\ee
Comparing this result with Eq.~(\ref{c_j^j}) we obtain again the equality
in Eq. (\ref{woot})
\begin{equation}
\label{s_j}
S_j(n)=\tau_j^{(n)}=\left\{
\ba{lcl}
0 &\ {\rm for}\ &\ n\le N \\
4s^2 c^{2(j-1)}\left[1-s^2 c^{2(j-1)}\right] &\ {\rm for}\ &\ j\le n\le N
\ea\right.
\end{equation}

To understand in more detail the meaning
of above expressions, let us consider the entanglement in the limit
$N\to\infty$ of a very large number of qubits in the reservoir.
We have to be careful with the definition of this limit. Let us first
recall the definition of homogenization. We want to obtain
homogenized qubits in states within some $\delta$-neighborhood of
the reservoir's state ($|0\r\l 0|$ in our case). In Section
\ref{sec5} we showed that if we have a large number of qubits $N$,
we can achieve an arbitrarily good homogenization, since in the
bound (\ref{4.4}) we can let $\delta\to 0$. In turn, the bound
(\ref{rel}) means that $\delta\to 0$ is obtained for $s\to 0$.
The behaviour of the expression $c^{2N}$ in this limit is as
follows: 
Since $s\to 0$, then $c\to 1$, but still $c<1$, therefore 
$\lim_{N\to\infty}c^{2N}=0$, too.
Now, looking at  Eqs.(\ref{8.18}),(\ref{8.19}),(\ref{s_0}) and
(\ref{s_j}), we see that in the limit $N\to\infty$ all the
concurrencies vanish. Therefore, the shared entanglement between
any pair of qubits is zero in this case, i.e. $\lim_{N\to\infty}
C_{jk}^{(N)} =0$. Also the entanglement shared between a given
qubit and rest of the homogenized system, expressed in terms of
the function $S_{k}{(N)}$ is zero: 
\be 
\lim_{N\to\infty} S_{k}{(N)} =
0; \qquad k=0,1,\dots N 
\ee
So,
that's how we define the limit $N\rightarrow\infty$: first, we
assume that the system qubit has interacted with all the $N$
qubits in the reservoir, for $N$ finite; then, we let $N$ go to
infinity, always assuming that we make the best possible
homogenization according to the bounds of Section \ref{sec5}.

\begin{figure}
\includegraphics[width=8cm]{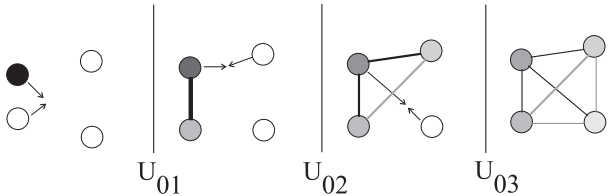}
\bigskip
\caption{
In this figure we schematically describe the process of
entanglement  between
the system qubit and the reservoir qubits via the homogenization.
The initial state of the whole system is shown
in the left part of the figure: We have the system qubit denoted by the
black circle, while the reservoir qubits are denoted as white
circles. After the interaction ($U_{01}$)
between the system and the first reservoir qubits a corresponding
change of states (represented in different degrees of the grey color)
and establishment of the entanglement (represented by the thick black
line) is exhibited. After the interaction ($U_{02}$) with the second
reservoir qubit, a three particle entangled state is created,
with various degrees of bi-partite entanglement (represented
by black and grey lines, where the grey line corresponds to the
entanglement between reservoir qubits which have not interacted
directly). In the right section of the figure we see the situation
after the interaction ($U_{03}$) of the system qubit with the third
reservoir qubit. All qubits are now entangled, black lines describe
the entanglement between the system and the reservoir qubits which is
established due to the direct interaction, while grey lines correspond
to the entanglement between reservoir qubits  induced by the interaction
with the system qubit.
}
\label{fig5}
\end{figure}

As a first result, it is instructive  to
realize that in the limit $N\rightarrow\infty$ (when
$\delta\rightarrow 0$) the functions $S_{0}(N)$ and $S_j(N)$ are
such that \be \lim_{N\rightarrow\infty} \frac{S_0(N)}{S_{k}(N)} =
1, \qquad j=1,\dots N \, , \ee which means that the entanglement
of the system qubit with the reservoir is the same as the
entanglement of an arbitrary reservoir qubit to the rest of the
homogenized system. This reflects the fact that not only states of
individual qubits are the same but also the amount of entanglement
between each of the qubits and the rest of the system are equal
(see Fig.~\ref{fig5}).

In spite of the fact that the pairwise entanglement between qubits
in the limit $N\rightarrow\infty$ tends to zero,
the information about the initial
state of the system qubit is distributed among the homogenized qubits.
Thus we have infinitely many infinitely small correlations between qubits
and it seems that the required information is lost. However, as $N$
goes to infinity we have infinitely many qubits and
the information  redistributed among them
has to be vanishingly small. If we sum up all the mutual concurrences between
all qubits we obtain a finite value
\be
\lim_{N\to\infty}\sum_{j<k}^N[C^{(N)}_{jk}]^2=\lim_{N\to\infty}
\frac{1}{2}\sum_{j=0}^N S_j(N)=2 .
\, ,
\label{6.26}
\ee
This supports our argument that the information about the initial state
of the system is ``hiden'' in mutual correlations between qubits of
the homogenized system. In the concluding section of the paper
we will study how this information can be recovered.

To clarify the meaning of Eq.~(\ref{6.26}), we recall the recent results
of Koashi {\it et al.} \cite{Koashi2000}. These authors have considered
a system of $N$ qubits composing the {\em web of entanglement}. That is,
each of $N(N-1)/2$ possible pairs of qubits is entangled, while the
degree of entanglement  is equal for all pairs. It has been shown that the
maximal degree of pairwise entanglement in the web of entanglement is
given by $C_{jk}^{(N)}=2/N$, that is the maximum tangle is $\tau=4/N^2$.
Given that there are $N(N-1)/2$ possible pairs we find that the total value
of the pairwise tangle is
\be
\lim_{N\rightarrow\infty}\frac{N(N-1)}{2} \frac{4}{N^2} = 2\, ,
\ee
which is the same value as found in the homogenized system under
consideration.

\section{Conclusions and discussion:\\ Applications of homogenization}
\label{sec7}

In this paper we have shown that one can choose a unitary transformation
that exchanges information between a system qubit and a qubit from
a reservoir, which, when applied sequentially to the system and each
qubit in the reservoir, will generate an evolution that has the
resevoir state as a fixed point. In fact, the state of the system qubit
and those of the reservoir qubits become  the
same. Moreover, this unitary transformation,
which we call the {\em partial swap} operation,  is
the {\em only} transformation, which is independent
of the initial states of the system and the  reservoir qubits, that
will accomplish this.

This result is interesting {\em per se}
since it allows us to understand in greater detail the dynamics
of open systems \cite{Davies}.
It is also a nontrivial fact that the partial swap operation applied
to the system qubit and a set of reservoir qubits allows us to
realize an arbitrary contractive map of the system qubit \cite{Preskill}.

On the other hand the results presented in the paper can be used in the
context of quantum information processing. Specifically,
quantum homogenization can be utilized  for quantum  cloning
and in a protocol realizing a {\em quantum safe with a classical
combination}.

\subsection{Quantum cloning}
It is well known that unknown quantum states cannot be copied perfectly.
 Specifically, Wootters and Zurek\cite{Wootters1982}
  have presented a very simple
proof that a perfect cloning transformation
for unknown quantum states is impossible. The ideal quantum cloning scenario
would look as follows:
The quantum cloner is initially prepared in a state
$|S\r$ which does not depend on the {\em unknown}
 state $|\psi\r$ of the input
qubit which is going to be cloned.
 In addition, a qubit onto which the
information is going to be copied  is available. This particle
is prepared in a {\em known}
state denoted as $|0\r$. The perfect copying
transformation ${\cal C}$ should be of the form
\be
|\psi\r |0\r|S\r
{\stackrel{\cal C}{\longrightarrow}} |\psi\r|\psi\r |S'\r
\label{5.1}
\ee
From the linearity of quantum mechanics it follows that the
cloning transformation (\ref{5.1}) is not possible.

Even though
 ideal cloning, i.e., the transformation (\ref{5.1}),
is prohibited by the laws  of quantum mechanics for an
{\em arbitrary} (unknown)
state $|\psi\r$, it has been shown that it still possible
to design quantum cloners which operate reasonably well
\cite{Buzek1996}. These quantum cloners
have been specified by the following conditions:

{\bf (i)} The  state of the original system and
its quantum copy at the
output of the quantum cloner, described by density operators
$\hat{\rho}^{(out)}_{1}$ and $\hat{\rho}^{(out)}_{2}$,
respectively, are identical, i.e.,
\begin{eqnarray}
\hat{\rho}^{(out)}_{1}   = \hat{\rho}^{(out)}_{2}.
\label{5.2}
\end{eqnarray}

{\bf (ii)} If no {\em a priori} information about the
{\em in}-state of the original system is available,
then it is reasonable to require
that {\em all} pure states should be copied equally
well. One way to implement
this assumption is to design a quantum
copier such that the distances between
density operators of each system at the output
$\hat{\rho}^{(out)}_{j}$
(where $j=1,2$) and the ideal density operator
$\hat{\rho}^{(id)}_j$
which describes the {\em in}-state of the original
mode are input state
independent.

{\bf (iii)}
Finally, it is also  required that
the copies are as close as possible to the ideal
output state, which is, of course, just the input state.
This means that the quantum copying transformation has
to minimize the distance between the output state $\hat{\rho}_{j}^{(out)}$
of the copied qubit and the ideal  state $\hat{\rho}_{j}^{(id)}$.

It has been shown by various authors that quantum cloners satisfying
the above conditions do exist \cite{Buzek1996,Gisin1997}.
Recently, experimental realizations
of these quantum machines have been reported as well
\cite{DeMartini2000,Huang2001}

However, this is not the only approach to quantum cloning;
one can formulate the problem from a slightly
different perspective using the ideas of quantum homogenization.
First, one can lift the condition (\ref{5.2}) that the
qubits at the output are in the same state, that is, it can be
assumed that the qubits at the output
  are in the states which are {\em similar},
but not identical. The second condition which might be lifted is
that the ``blank'' qubit is initially in the {\em known} state $|0\r$.
We can instead
assume that both the input state of the original  and that
of the ``blank'' are unknown.
If this point of view is adopted then the quantum homogenization
as characterized by the conditions (\ref{1.3}) and (\ref{1.4}) can
be successfully used for approximate cloning. Specifically, in this
scenario the reservoir qubits play the role of originals, that is,
it is the state $\xi$ we want to copy, while  the system, which
is supposed to be homogenized, (this system is initially in an unknown
state  $\varrho_S^{(0)}$) plays the role of the ``blank'' system onto
which the information is going to be copied.
From the description of quantum homogenization we
see that quantum cloning in this context is a process in which
we start with $N$ reservoir qubits, all in the same state $\xi$,
and we end up with $N+1$ qubits in states which are very close
(how close depends on the value of $N$) to the state $\xi$,
so we have performed a version of $N\rightarrow N+1$ cloning
on the reservoir state.

\subsection{Quantum safe with a classical combination}

After the system qubit is homogenizaed
it is in the same state as the reservoir qubits,
so we can ask: What happened
to the information encoded in the initial state of the system qubit?
Is it irreversibly lost? Certainly not, because
we considered only unitary transformations,
and that means that the information encoded in the initial state
of the system qubit is not lost but it is
transferred into  quantum correlations between all of the qubits.
The parameters
characterizing the state of the system are transformed into parameters
determining the entanglement shared among the system and reservoir qubits.
One question is whether the initial state of the system qubit can be
recovered.

The process of homogenization is described by a sequence of unitary
operations. Consequently, it can be reversed: That is the
homogenized system can be ``unwound'' and the original state
of the system  $\varrho_S^{(0)}$ and the reservoir $\xi$ can be recovered.
Perfect unwinding can be performed only when the $N+1$ qubits
of the output state interact, via the inverse of the original
partial-swap operation, in the ``correct'' order. The system
particle must be identified from among the $N+1$ output qubits,
and it and the reservoir particles must interact in the reverse
of the order in which they originally interacted. Therefore, in
order to unwind the homogenized system, the classical
information about the ordering of the particles is vital. Obviously,
if there are at the output $N+1$ particles, then there exists $(N+1)!$
permutations of possible orderings, only one of which will reverse
the original process.  The probability to choose the system particle,
which is in the state $\varrho_S^{(N)}$, correctly is $1/(N+1)$.
Even when this particle is chosen succesfully, then
there are still $N!$ different possibilities in choosing the sequence
of interactions with the reservoir qubits.  If one
has no knowledge about the output particles, the probability of
successfully unwinding the homogenization transformation is
$1/(N+1)!$.  As we shall see, if at the beginning of the unwinding
process the reservoir particle is chosen incorrectly then the
whole process leads to a completely wrong result.

Therefore we can consider the quantum homogenization as a process
that generates a combination to a quantum safe.  The combination
is the sequence in which the reservoir particles interact with the
system particles, and the object in the safe is the initial state
of the system particle.  The combination consists of classical
information, and the object in the safe consists of quantum
information.  The security here is given by the homogeneity of the
final ensemble; it is difficult to distinguish among the particles
by measuring them.  The unwinding process can be performed
reliably only when the combination is available. An important
aspect of this scheme is that if one has tried one possible
unwinding of the state, and measured the result to gain some
information about it, it is not possible at that point to try to
unwind it in a different way. That is, the nature of quantum
mechanical measurement prevents repeated unwinding procedures on
the same homogenized set of particles.

\begin{figure}
\includegraphics[width=8cm]{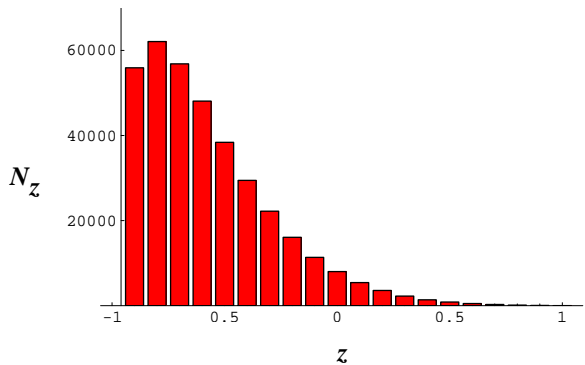}
\bigskip
\caption{The result of the unwinding process with a trial-and-error
strategy, when the system qubit is correctly chosen from a set of
10 homogenized particles.
We plot a histogram representing number of reconstructed
states of the system qubit
with $s$ falling into the bin with $z=z_{n}\pm 0.05$.
There are altogether $9!$ sequences we have checked and
just one results in a correct reversal of the homogenization
process.
}
\label{fig3}
\end{figure}

To illustrate the above protocol let us assume that we begin with
the system qubit in the state $|1\r$
and nine reservoir qubits in the $|0\r$.
After quantum homogenization we try, randomly, to unwind the
process. Let us assume that we are lucky and
we have chosen  the first qubit in the unwinding process
correctly, that is, it is the original system qubit.
Even with this good start,  we have to find the rest of the combination,
the proper sequence of the reservoir qubits,
in order to completely ``open'' the quantum safe. Here we adopt a
trial-and-error strategy, and we test all possible permutations of the
reservoir qubits.
Obviously,
in this case just {\em one} sequence is correct, i.e. will result
in opening the quantum safe and recovering the system state.
All $9\, != 362 880$ possible permutations of the reservoir
qubits which were tested.
Since we have chosen the states of the system and the reservoir qubits
to be two orthogonal basis states of a single qubit,
we can parameterize
the reconstructed system state with just a single parameter $s$, i.e.
$\rho_{unwound}=\frac{1}{2}(\openone+ z \cdot \sigma_z)=
\frac{1+z}{2}|0\r\l 0|
+\frac{1-z}{2}|1\r\l 1|$, such that $-1\leq z\leq 1$. In Fig.~
\ref{fig3}
we plot the histogram representing the number $N_s$ of reconstructed
states of the system qubit
with $s$ falling into the bin with $z=z_{n}\pm 0.05$.
We see that a randomly chosen combination will not
open the quantum safe.  In fact, most of the
reconstructed states are within the interval $-1\leq z \leq 0$,
i.e. between the reservoir state and the completely random state.

\begin{figure}
\includegraphics[width=8cm]{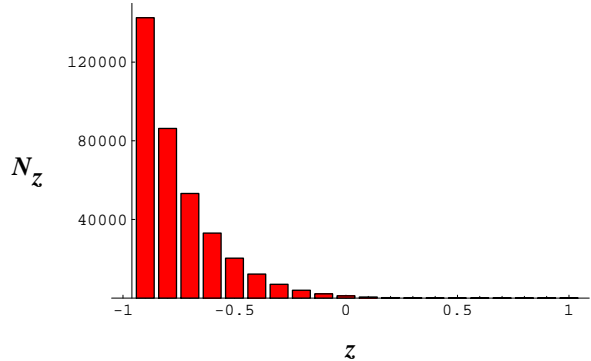}
\bigskip
\caption{The result of the unwinding process with a trial-and-error
strategy, when the system qubit is chosen incorrectly from a set of
10 homogenized particles. In the figure we represent results of
$9\cdot 9!$ random unwindings. None of these sequences lead to the
correct reversal of the original homogenization.
We plot the histogram representing number of reconstructed
states of the system qubit
with $s$ falling into the bin with $z=z_{n}\pm 0.05$.
}
\label{fig4}
\end{figure}

Let us now consider what happens when we choose the wrong qubit as
the system qubit, i.e. what we have chosed as the system qubit
was, in fact, one of the
original reservoir qubits. As can be checked explicitly, in this
case there is no way to correctly unwind the homogenization process.
Obviously, with no prior knowledge, the probability
to chose an incorrect system qubit from a set of $N+1$ homogenized
qubits is $N$ time larger than the probability to chose the system
qubit correctly. In addition, there are $N\cdot N!$ different
sequences for the unwinding
procedure in this case and none of them results in the initial state.
In Fig.~\ref{fig4} we plot the results of these unwinding procedures
for the same choice of the initial states as in the previous case.

We can conclude that the process of quantum homogenization can
be unwound (i.e. {\em reversed}) if and only if classical
information about the order of reservoir qubits is available.
If this information is discarded, the process becomes
irreversible even though the overall dynamics is unitary.  This
irreversibility can be used to protect quantum information.
A detailed analysis of the security of the protocol we have
proposed to do this remains to be done, but the example we have
treated numerically strongly suggests that quantum information
protected in this way is very secure.

\acknowledgements
This was work supported in part
by  the European Union projects EQUIP (IST-1999-11053),
QUBITS (IST-1999-13021),
by the  National Science Foundation under grant
PHY-9970507, and by  the Slovak Academy of Sciences.
N.G. and V.S. acknowledge partial
financial support from the Swiss FNRS and the Swiss OFES within
the European project EQUIP (IST-1999-11053).

\end{multicols}


\begin{thebibliography}{99}
\bibitem{footnote1}
We implicitely assume that the reservoir qubits are
distinguishable.  The validity of this
assumption depends on the physical realization of the qubit. For
instance, if the qubits are nuclear spins --- or more generally,
if each qubit is a degree of freedom of a given atom --- the
assumption is valid, since atoms are distinguishable from one
another under normal conditions.
\bibitem{note0}
This is very easy to see: suppose that $\xi$ and $\xi'$ are two
fixed points: then they must satisfy $D(\xi,\xi')\leq
k\,D(\xi,\xi')$ for a given $k\in[0,1[$. This is impossible unless
$D(\xi,\xi')=0$.
\end{thebibliography}
\end{document}